\documentclass{PoS}

\usepackage{upgreek}
\usepackage{multirow}
\usepackage[mathlines]{lineno}

\title{Measurements of open charm hadron production in Au+Au Collisions at $\sqrt{s_{\rm{NN}}}$ = 200 GeV at STAR}

\ShortTitle{Open charm hadron production at STAR}

\author{\speaker{Guannan Xie} (for the STAR Collaboration)\\
        University of Illinois at Chicago, Chicago, IL 60607, USA\\
        E-mail: \email{xieguannanpp@gmail.com}}

\abstract{
  We report on the measurements of production of various charmed hadrons in Au+Au collisions at $\sqrt{s_{\rm{NN}}}$ = 200 GeV (including $D^{0}(\overline{D^{0}})$ and $\Lambda_{c}^{\pm}$) obtained via topological reconstruction, utilizing the Heavy Flavor Tracker at STAR. Precise results on the $D^{0}$ yields from the 2014 data are reported for a wide transverse momentum range down to 0 in various centrality bins. With the high-statistics data collected in 2014 and 2016, and the usage of a supervised machine learning algorithm for signal-to-background separation, the first measurement of the centrality and transverse momentum dependences of $\Lambda_{c}^{\pm}$ production is shown. Finally, the total charm quark cross section extracted from these measurements in Au+Au collisions at $\sqrt{s_{\rm{NN}}}$ = 200 GeV is presented.
}

\FullConference{International Conference on Hard and Electromagnetic Probes of High-Energy Nuclear Collisions\\
		30 September - 5 October 2018\\
		Aix-Les-Bains, Savoie, France}

\begin{document}

\section{Introduction}

Because of their large mass, heavy quarks (charm and bottom) are predominately created through initial hard scatterings in heavy-ion collisions at RHIC and the LHC. They experience the whole evolution of the system and thus are suggested to be an important tool for studying the properties of the Quark Gluon Plasma (QGP) produced in heavy-ion collisions~\cite{StarWhitePaper,LhcPaper}. The modification of their production in transverse momentum ($p_{T}$) due to energy loss and in azimuth due to anisotropic flows is sensitive to heavy-quark dynamics in the partonic QGP phase. Recently, measurements at RHIC and the LHC have indicated strong energy loss and large elliptic flow for open charm hadrons, similar in magnitude to those of light-flavor hadrons~\cite{Star_D_RAA,Star_D_v2,Alice_D}. The observed enhancements of $\Lambda_{c}^{\pm}$ and $D_{s}^{\pm}$ production yields in Au+Au collisions suggest that the coalescence mechanism also plays an important role for charm quark hadronization. Study of the charm quark hadronization mechanism in the QGP is also crucial for understanding the charm meson suppression in heavy-ion collisions and charm quark energy loss in QGP~\cite{Cacciari}.

In these proceedings, we report on the production of various charmed hadrons in Au+Au collisions at $\sqrt{s_{\rm{NN}}}$ = 200 GeV obtained via topological reconstruction, utilizing the Heavy Flavor Tracker (HFT) at STAR. Precise measurements of the $D^{0}$ yields are reported for a wide $p_T$ range down to 0. The $D^{0}$ $R_{\rm AA}$ and $D^{0}$ $R_{\rm CP}$ are also reported in various centrality bins and compared to those of light-flavor hadrons as well as model calculations. The first measurement of the centrality and $p_T$ dependences of the $\Lambda_{c}^{\pm}$/$D^{0}$ ratio in heavy-ion collisions is shown. Finally, the total charm quark cross section extracted from these measurements in Au+Au collisions is presented.

\section{Experimental and Analysis}

The STAR experiment at RHIC is a large-acceptance detector covering full azimuth and pseudorapidity of $|\eta| < 1$. Data taken by the STAR experiment with the HFT installed were used for this analysis. The HFT consists of four sub-detectors, two layers of Pixel detectors (PXL) close to the beam pipe, the Intermediate Silicon Tracker (IST) and the Silicon Strip Detector (SSD) at the outermost layer. The HFT is a high resolution silicon detector which provides a track pointing resolution of less than 50 $\upmu \textup{m}$ for kaons with $p_T$ = 750 MeV/$c$. The excellent pointing resolution significantly improves the signal-to-background ratio by reducing the combinatorial background when topologically reconstructing charm hadrons decaying close to the collision vertex, especially in heavy-ion collisions. The particle identification ($\pi, K, p$) is performed by measuring the ionization energy loss (dE/dx) in the Time Projection Chamber (TPC) and velocity using the Time-Of-Flight (TOF) detector. In total, about 900 million minimum-bias (MB) Au+Au events from the year 2014 and 1 billion events from the year 2016 were used for this analysis. These events were required to have primary vertices (PV) within 6 cm from the center of the STAR detector along the beam direction to ensure uniform HFT acceptance.

The $D^0$ mesons were reconstructed via the hadronic decay channel: $D^0\rightarrow K^-\pi^+$ ($B.R.\sim$ 3.89\%) and the $\Lambda_{\textup{c}}^+$ baryons through $\Lambda_{\textup{c}}^+\rightarrow p^+K^-\pi^+$ ($B.R.\sim$ 6.23\%), and their charge conjugate channels. In order to reach the high pointing precision, all the daughter tracks are required to have at least three hits in the HFT sub-detectors with two of them in the two PXL layers.
The cuts on the topological variables for this analysis are optimized using a Toolkit for Multivariate Data Analysis (TMVA) package integrated in the ROOT framework in order to obtain the highest signal significance~\cite{TMVA}. The Rectangular Cut optimization method from the TMVA package is chosen for $D^0$, and a supervised learning algorithm, Boosted Decision Trees (BDT) from the Toolkit is used for $\Lambda_{\textup{c}}^\pm$ signal and background separation. The TPC acceptance and tracking efficiency are obtained using the standard STAR TPC embedding technique. The particle identification efficiency and the HFT acceptance and tracking plus topological cut efficiency are obtained using a data-driven simulation method in order to fully capture the real-time detector performance.

\section{Physics Results}

\begin{figure}[htbp]
\hspace{+0.5cm}
\begin{minipage}[b]{0.45\linewidth}
\begin{center}
\includegraphics[width=\textwidth]{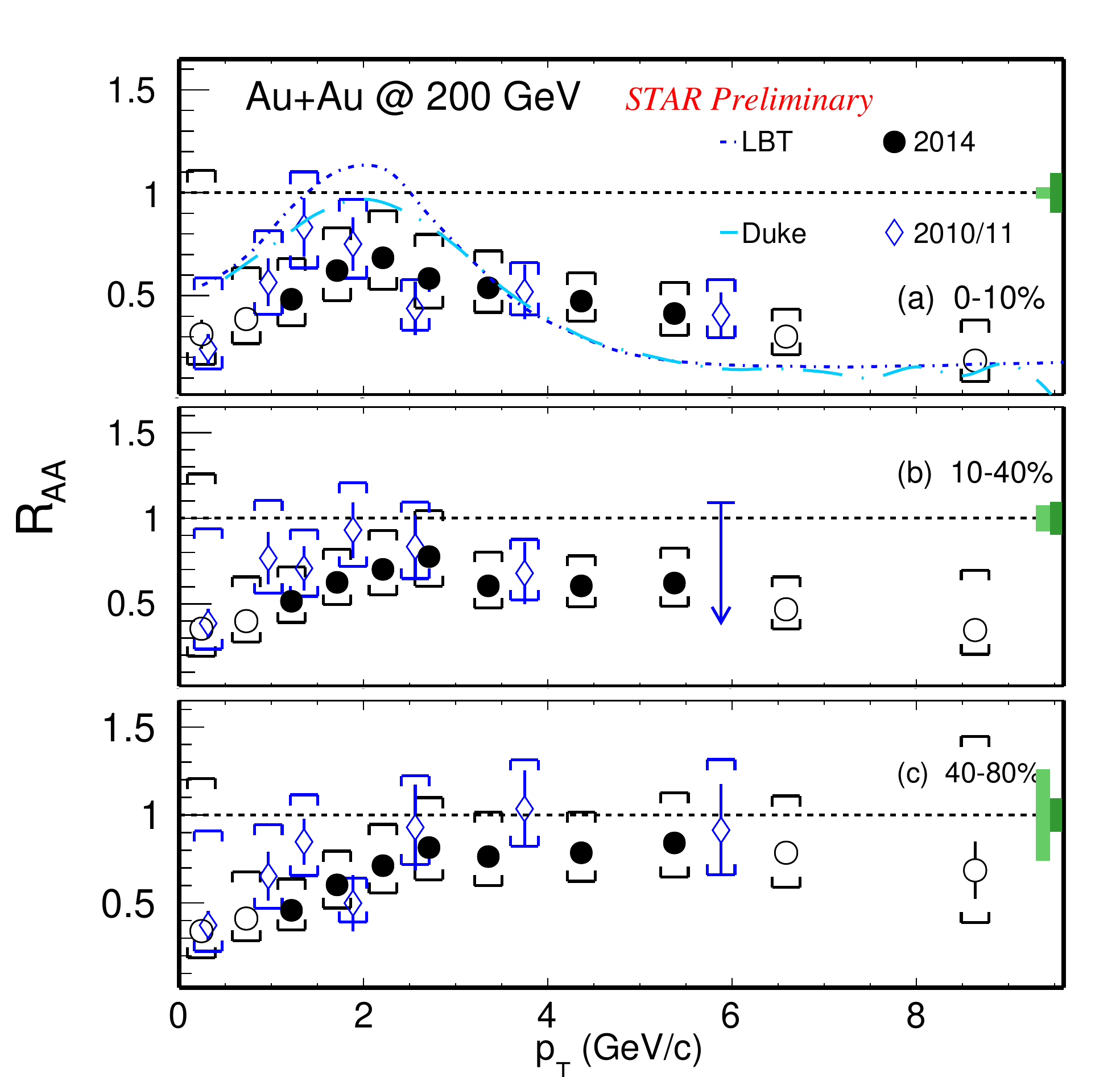}
\end{center}
\end{minipage}
\hspace{+0.5cm}
\begin{minipage}[b]{0.45\linewidth}
\begin{center}
\includegraphics[width=\textwidth]{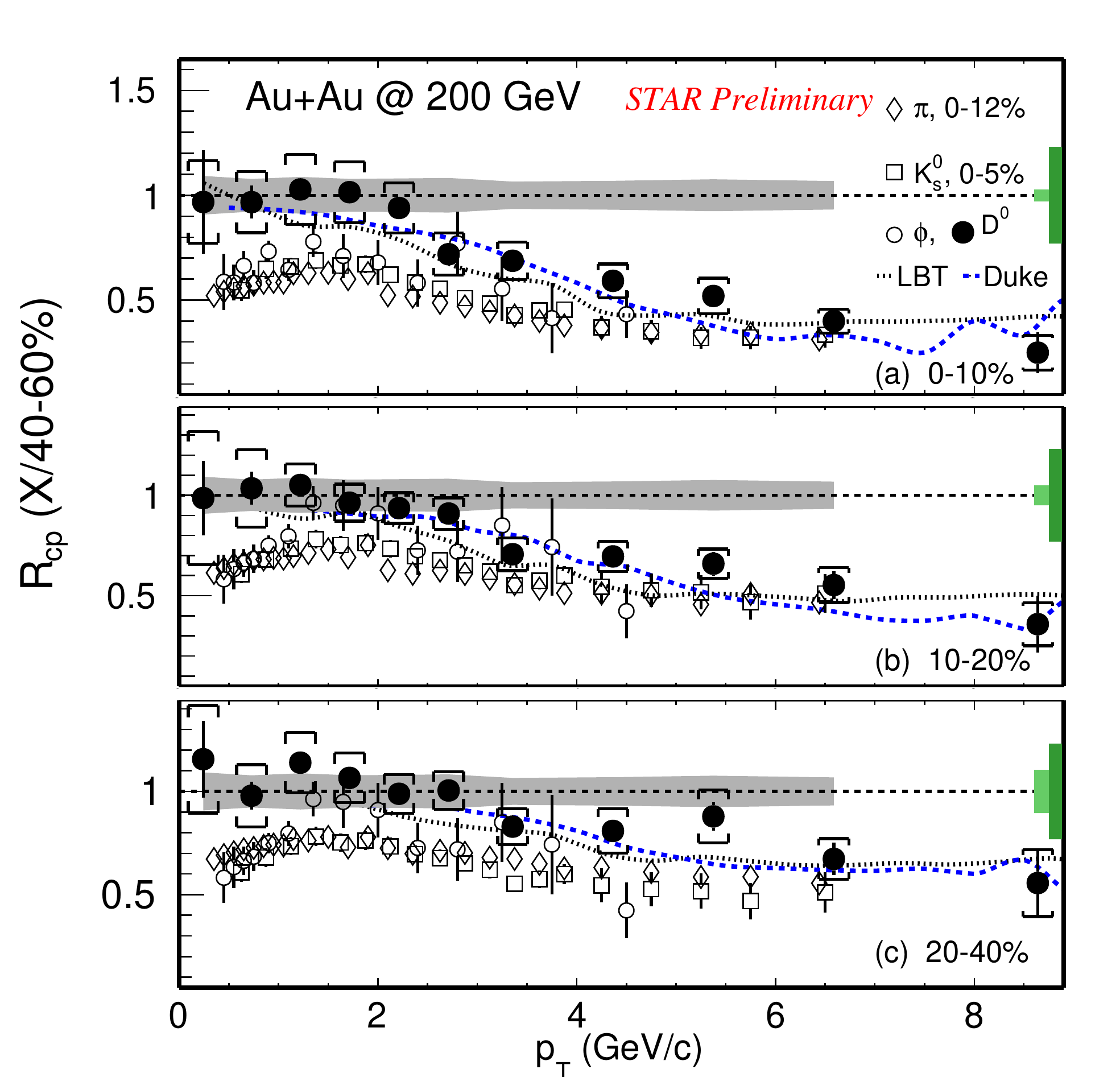}
\end{center}
\end{minipage}
\caption{ (Left) $D^{0}$ $R_{\rm AA}$ in Au+Au collisions at $\sqrt{s_{\rm{NN}}}$ = 200 GeV for 0--10\% (a), 10--40\% (b) and 40--80\% (c) centrality bins, respectively. (Right) $D^{0}$ $R_{\rm CP}$ with the 40--60\% spectrum as the reference for different centrality classes in Au+Au collisions compared to those of light-flavor and strange mesons ($\pi^{\pm}$, $K^0_{S}$ and $\phi$). 
The grey bands around unity depict the systematic uncertainty due to vertex resolution correction. The light and dark green boxes represent the global uncertainties in the spectra for the given centrality and the reference, respectively.}
\label{fig:D0_RAA_Rcp}
\end{figure}

Figure~\ref{fig:D0_RAA_Rcp} left panel shows the $D^0$ $R_{\rm AA}$ with the $p$+$p$ measurement~\cite{Star_D_pp} as the reference for different centrality bins 0--10\% (a), 10--40\% (b) and 40--80\% (c), respectively. The new $R_{\rm AA}$ measurements are also compared to the previously published results using only the STAR TPC after recent correction~\cite{Star_D_RAA}. The $p$+$p$ $D^0$ reference spectrum is updated using the latest global analysis of charm fragmentation ratios from Ref.~\cite{charm_frag} and also by taking into account the $p_T$ dependence of the fragmentation ratio between $D^0$ and $D^{*\pm}$ from PYTHIA. The new measurement with the HFT detector shows a nice agreement with the previous measurement without the HFT. The brackets on the data points depict the total systematic uncertainty dominated by the uncertainty in the $p$+$p$ reference spectrum. The open circles of the first two and last two data points indicate that those are calculated with an extrapolated $p$+$p$ reference. From low to intermediate $p_{T}$ region, the $D^0$ $R_{\rm AA}$ shows a characteristic structure that is qualitatively consistent with the expectation from model predictions that charm quarks gain sizable collective motion during the medium evolution. The large uncertainty in the $p$+$p$ baseline needs to be further reduced before making more quantitative conclusions. The right panel of Fig.~\ref{fig:D0_RAA_Rcp} shows the $D^0$ $R_{\rm CP}$ for different centralities as a function of $p_{T}$ with the 40--60\% centrality spectrum as the reference. As a comparison, $R_{\rm CP}$ of charged pions, $K_{s}^{0}$ and $\phi$ in the corresponding centralities are also plotted in each panel. The measured $D^0$ $R_{\rm CP}$ in central 0--10\% collisions shows a significant suppression at $p_{T}>$ 5\,GeV/$c$. The suppression level is similar to that of light-flavor and strange mesons and the suppression gradually decreases when moving from central collisions to mid-central and peripheral collisions. The $D^0$ $R_{\rm CP}$ for $p_{T}$\,$<$\,4\,GeV/$c$ does not show a modification with centrality, in contrast to light-flavor hadrons. Calculations from the Duke group and the Linearized Boltzmann Transport (LBT) are also compared to the data~\cite{duke,lbt}. Both collisional and radiative energy losses are included in these two calculations, and the parameters used in the models are tuned to reproduce the previously published results~\cite{Star_D_RAA}. Both model calculations match our new measured data well while the improved precision of the new measurements is expected to further constrain the theoretical model calculations.

\begin{figure}[htbp]
\hspace{+0.5cm}
\begin{center}
\includegraphics[width=0.8\textwidth]{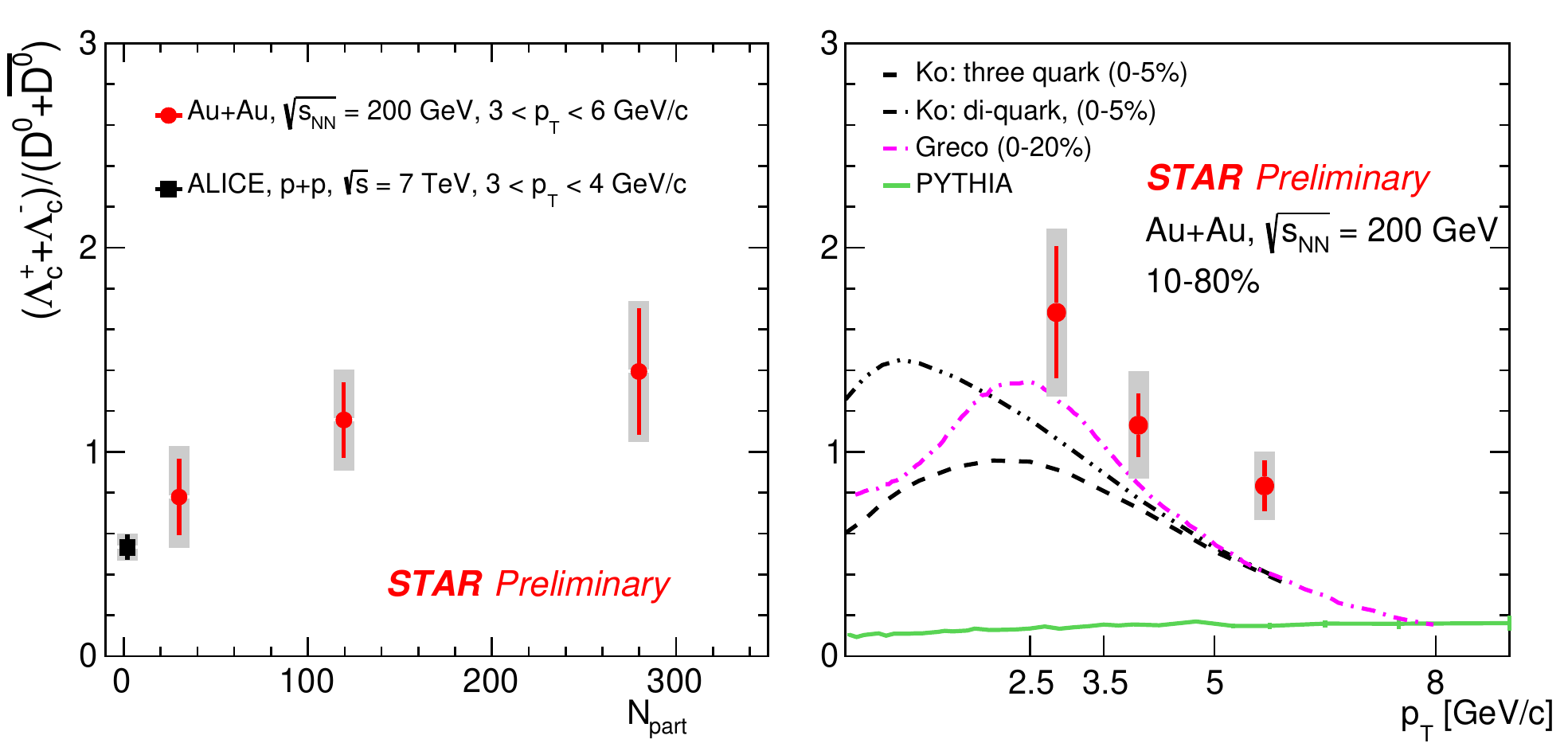}
\end{center}
  \caption{ (Left) $(\Lambda_c^++\Lambda_c^-)$/$(D^0+\overline{D^0})$ ratio as a function of $N_{\rm part}$ in $3 < p_{T} < 6$\,GeV/$c$. (Right) $(\Lambda_c^++\Lambda_c^-)$/$(D^0+\overline{D^0})$ ratio as a function of $p_T$ for the 10--80\% centrality class.}
\label{fig:Lc_pt_cent}
\end{figure}

Figure~\ref{fig:Lc_pt_cent} shows in the left panel the measured $(\Lambda_c^++\Lambda_c^-)$/$(D^0+\overline{D^0})$ ratio as a function of $N_{\rm part}$ in $3 < p_{T} < 6$\,GeV/$c$. There is a clear increasing trend towards more central collisions while the value in the peripheral collisions is comparable with the measurement in $p$+$p$ collisions at 7 TeV from ALICE~\cite{aliceLc}. The right panel shows the $(\Lambda_c^++\Lambda_c^-)$/$(D^0+\overline{D^0})$ ratio as a function of $p_T$ for the 10--80\% centrality class. The values show a significant enhancement compared to the calculations from PYTHIA. The enhancement is also larger than the statistical hadronization model (SHM) prediction~\cite{shm}. In Ko's~\cite{Ko} and Greco's model calculations~\cite{greco}, which include coalescence hadronization of charm quarks, the predicted ratio is comparable to our measurement, but tends to underestimate the data for high $p_T$. However, one needs measurements at low $p_T$ to further differentiate between three-quark and di-quark recombination scenarios.

\begin{table*}
\centering{
\caption{Total charm cross-section per binary nucleon collision at midrapidity in Au+Au and $p$+$p$ collisions at 200 GeV.}
\begin{tabular}{c|c|c} \hline \hline
  \multicolumn{2}{c|} {Charm Hadron}   & Cross Section d$\sigma$/dy($\upmu$b) \\ \hline
  \multirow{4}{*}{Au+Au } & \hspace{1cm} $D^0$ \hspace{1cm}  & 41 $\pm$ 1 (stat) $\pm$ 5 (sys) \\ \cline{2-3} 
  & \hspace{1cm} $D^+$ \hspace{1cm} & 18 $\pm$ 1 (stat) $\pm$ 3 (sys)\\ \cline{2-3}
  \multirow{2}{*}{(10-40\%)} & \hspace{1cm} $D_s^+$ \hspace{1cm} & 15 $\pm$ 1 (stat) $\pm$ 5 (sys)\\ \cline{2-3}
  & \hspace{1cm} $\Lambda_c^+$ \hspace{1cm} & 78 $\pm$ 13 (stat) $\pm$ 28 (sys)\\ \cline{2-3}
  & \hspace{1cm} total $c\overline{c}$ \hspace{1cm} & 152 $\pm$ 13 (stat) $\pm$ 29 (sys)\\ \hline 
  $p$+$p$ & \hspace{1cm} total $c\overline{c}$ \hspace{1cm} & 130 $\pm$ 30 (stat) $\pm$ 26 (sys)\\ \hline \hline
\end{tabular}
}
\label{table:crossX}
\end{table*}

Besides the $D^0$ and $\Lambda_c^{\pm}$, STAR also has performed measurements of $D^{\pm}$ and $D_s^{\pm}$~\cite{STAR_Dpm,STAR_Ds} in Au+Au collisions at $\sqrt{s_{\rm{NN}}}$ = 200 GeV. With these various charmed hadron measurements, the total charm quark cross section per binary nucleon collision was obtained and listed in Table 1. For the $D^0$, the measurements were performed down to $p_T = 0$ while the other charmed hadrons are extrapolated to low $p_T$ which results in sizable uncertainties for these cross section measurements. The total $c\overline{c}$ cross section per binary nucleon collision in Au+Au collisions is consistent with that in $p$+$p$ collisions within uncertainties. However, the charm hadrochemistry is modified in heavy-ion collisions compared to that in $p$+$p$ collisions.

\section{Summary}

We have presented the recent measurements of production of various charmed hadrons in Au+Au collisions at $\sqrt{s_{\rm{NN}}}$ = 200 GeV obtained by the STAR experiment at RHIC. The measured $D^0$ $R_{\rm CP}$, $R_{\rm AA}$ and $(\Lambda_c^++\Lambda_c^-)$/$(D^0+\overline{D^0})$ ratio are presented and compared to model calculations.

\end{document}